\documentclass[preprint,showpacs,preprintnumbers,amsmath,amssymb]{revtex4}

\usepackage{graphicx}
\usepackage{dcolumn}
\usepackage{bm}

\begin{document}


\title{Magnetic induction dependence of the dispersion of magnetoplasmon in a two-dimensional electron gas with finite layer thickness}

\author{T. Uchida$^{(b)}$}
\author{N. Hiraiwa$^{(a)}$}
\author{K. Yamada$^{(b)}$}
\author{M. Fujita$^{(b)}$}
\author{T. Toyoda$^{(b)}$}
 \email{toyoda@keyaki.cc.u-tokai.ac.jp}
\affiliation{$^{(a)}$SHINANO FUJITSU LTD.,
Oaza-Nosakada 935, Iiyama, Nagano 389-2233, Japan \\
$^{(b)}$Dept. of Physics, Tokai University, Kitakaname, Hiratsuka, 
Kanagawa 259-1292, Japan}

\date{\today}

\begin{abstract}
Magnetic induction dependence of the dispersion of longitudinal magnetoplasmon in a two-dimensional electron gas with finite layer thickness under a static uniform magnetic field normal to the layer plane is calculated using the self-consistent linear response approximation.  Two longitudinal magnetoplasmon modes are obtained. The calculated dispersion agrees with the experiment by Batke {\it et al.} [Phys. Rev. B {\bf 34}, 6951 (1986)].
\end{abstract}

\maketitle

\section{Introduction}
The dispersion of plasmons in a two-dimensional electron gas (2DEG) has been studied theoretically by many authors for decades \cite{{Quinn-Ferrel},{Ferrell},{Stern1967},{Zhu1988},{Eliasson1987},{Hawrylak1988}}. 
The nature of plasmons changes drastically when a strong uniform magnetic field normal to the 2DEG layer is applied, because the motion of the electrons in the layer is completely quantized to form Landau orbitals. 
In 1969, Greene, Lee, Quinn, and Rodoriguez \cite{Greene69} developed a linear response theory for a degenerate three-dimensional electron gas in such a strong magnetic field. They calculated the current response functions of the electrons and obtained dispersion relations of magnetoplasmons in the three-dimensional system. Their theory was applied to a 2DEG by Chiu and Quinn \cite{Chiu74}. Horing and Yildiz \cite{{Horing73},{Horing76}} also developed a quantum theory of longitudinal dielectric response properties of a 2DEG in a magnetic field. Zhang and Gumbs \cite{Zhang93} calculated the correlations and local field corrections for 2DEG in a strong magnetic field and also obtained dispersion relations of  longitudinal magnetoplasmons. Kallin and Halperin \cite{Kallin} assumed a filled Landau level and used the Bethe-Salpeter equation to calculate the electromagnetic response function of a 2DEG in a magnetic field. Magnetoplasmon excitations from partially filled Landau levels were calculated by MacDonald, Oji, and Girvin \cite{MacDonald85}. The Hofstadter energy spectrum in far-infrared absorption was studied by Gudmundsson and Gerhardts \cite{Vidar96}. 

Recent experimental discovery of the existence of plateaus in the filling factor dependence of the dispersion of magnetoplasmon \cite{Holland04} has revived the significance of the theoretical work by by Chiu and Quinn \cite{Chiu74} and by Horing and Yildiz \cite{{Horing73},{Horing76}}, which provided a microscopic basis for the semiclassical dispersion formula \cite{Toyoda2008}. Under these circumstances it is necessary to develop Chiu-Quinn-Horing-Yildiz theory further. 

The aim of this paper is to advance their theory by carrying out fully analytical calculation of the dispersion, i.e., without depending on numerical computations, including the effects of the finite layer thickness on the dispersion relation of the longitudinal plasmon in a 2DEG embedded in a bulk dielectric and subject to a quantizing magnetic field, within the theoretical framework of the SCLRA used in our previous work \cite{{Toyoda1984},{Toyoda1998},{FHMT2007},{Fukuda2004},{Fukuda2005}}. 
We also present the calculation of the retarded density-density response function on the basis of the canonical commutators of the electron field operators in the Heisenberg picture in details. The calculation becomes much more transparent and simpler than the previous ones \cite{{Greene69}, {Chiu74},{Horing73},{Horing76}}, and can be readily applied to other models such as the zero-mass Dirac model for graphene straightforwardly.   

In the next section we illustrate the basic idea of SCLRA. In Sec. III, we present the quantum field theoretical calculation of the retarded density-density response function of a 2DEG in a magnetic field. In Sec. IV, we derive the SCLRA equation to determine the dispersion relations of the magnetoplasmons in the long-wavelength limit. In Sec. V, we calculate the magnetic induction dependence of the dispersion, and compare it with the experiment by Batke {\it et al} \cite{Batke86} to examine the validity of the theory. In Sec. VI, we give concluding remarks.

\section{Self-consistent linear response approximation}
In the Coulomb gauge condition, which we adopt in this work, the classical electromagnetic fields are given in terms of the transverse vector potential and the scalar potential. The scalar potential satisfies Poisson's equation
\begin{align}
\nabla^2 A_0 ({\bm x},t)= 4\pi e \epsilon^{-1} \rho({\bm x},t), 
\label{Poisson}
\end{align}
where $-e$ is the electron charge, $\epsilon$ is the dielectric constant, and $\rho$ is the electron number density. We choose a Cartesian coordinate system with the $x_3$ axis parallel to the magnetic field and the $x_1-x_2$ plane corresponding to the electron layer. Throughout this paper we use the notation ${\bm x}\equiv(x_1,x_2,x_3)$ and $\nabla\equiv (\partial_1,\partial_2,\partial_3)$ for the three-dimensional Cartesian coordinates. If the charge density is given and a proper boundary condition is specified, this Poisson's equation may be solved. Under the Coulomb gauge condition, the scalar potential corresponds to an instantaneous interaction between charged particles and hence there is no retardation effect in it. The basic assumption of the classical theory is that the space-time distribution of the charge density is a given quantity and, therefore, would not be affected by the electric field. 
In the self-consistent linear response approximation (SCLRA), which has been widely used for half a century in quantum many-body theory under various names such as random phase approximation \cite{{Quinn-Ferrel}}, we assume the existence of the interactively induced scalar potential fluctuation $A_0^{(1)}$ and electron charge density fluctuation $\delta\rho$. Then the scalar potential can be split into two parts,
$ A_0 = A_0^{(0)} + A_0^{(1)} $
and the electron density can also be written as
$ \rho = \rho^{(0)} + \delta\rho $
with
\begin{align}
\nabla^2 A_0^{(1)} ({\bm x},t)= 4\pi e \epsilon^{-1} \delta\rho({\bm x},t).
\label{Poisson2}
\end{align} 
To take into account of the effects of the scalar potential $A_0^{(1)}$ on the dynamics of the electron gas, we use the linear response approximation
\begin{align}
\delta\rho({\bm x},t) 
= 
\frac{-e}{\hbar}\int dt' \int d^3{\bm x}'  
{\mathcal D}^{(3D)}({\bm x},t;{\bm x}',t') A_0^{(1)}({\bm x}',t'). 
\label{linear3D}
\end{align}
Here the dynamics of the electrons is contained in the electron retarded density-density response function defined as
\begin{align}
{\mathcal D}^{(3D)}({\bm x},t;{\bm x}',t') = -i\theta(t-t') 
\sum_{\alpha \beta} 
\left< \left[ 
\Psi_\alpha^\dagger({\bm x},t)  \Psi_\alpha({\bm x},t),  
\Psi_\beta^\dagger({\bm x}',t')  \Psi_\beta({\bm x}',t')
\right] \right> .
\label{3dimD}
\end{align}  
Here $\Psi_\alpha({\bm x},t)$ and $\Psi_\beta^\dagger({\bm x},t)$ are the second quantized Schr{\"o}dinger field operators describing the electrons and satisfy the equal-time canonical anticommutation relation
\begin{align}
\left\{ \Psi_\alpha({\bm x},t), \Psi_\beta^\dagger({\bm x}',t)  \right\} = 
\delta_{\alpha\beta}\delta({\bm x}-{\bm x}') ,
\end{align}
where we have defined the anticommutator product, $\{A, B\}\equiv AB+BA$.
The function $\theta$ is defined as 
$ \theta(\tau) = 0 $ for $\tau<0$ and $ \theta(\tau) = 1 $ for $\tau>0$. 
The Greek subscripts denote the spin variables. The bracket $<...>$ denotes the grand canonical ensemble expectation value in an equilibrium mixed state.

\section{Retarded density-density response function}
The dynamics of these field operators  $\Psi_\alpha({\bm x},t)$ and $\Psi_\beta^\dagger({\bm x},t)$ is determined by the Hamiltonian 
\begin{align}
H=\sum_\alpha \int d^3{\bm x} \Psi_\alpha^\dagger({\bm x},t) \xi_A(\partial) \Psi_\alpha({\bm x},t) + H_{spin} .
\label{hamiltonian-1}
\end{align}
The second term on the right-hand side, $H_{spin}$, is the interaction between the electron spin and the magnetic field 
\begin{align}
H_{spin}=\sum_\alpha \frac{g}{2}\mu_B \sigma (\alpha) 
\int d^3{\bm x} B 
\Psi_\alpha^\dagger({\bm x},t)\Psi_\alpha({\bm x},t) ,
\end{align}
where $g$ is the effective $g$-factor, $\mu_B$ is Bohr magneton, $\sigma(\uparrow)=1$, and $\sigma(\downarrow)=-1$. Note that Bohr magneton $\mu_B= e\hbar/2m_0 c$ contains the electron rest mass $m_0$.
The electron single-particle energy operator $\xi_A(\partial)$ is defined as
\begin{align}
\xi_A(\partial)=\frac{1}{2m}\sum_{k=1}^3
\left( -i\hbar\partial_k + ec^{-1}A_k({\bm x}) \right)^2 - \mu_0 ,
\end{align}
where $m$ is the electron effective mass and $\mu_0$ is the chemical potential. In order to describe the static uniform magnetic field normal to the $x_1$-$x_2$ plane we adopt the vector potential in the form $A_1=-Bx_2$, $A_2=A_3=0$. 
We assume that the electrons are confined in the $x_1$-$x_2$ plane by a confining potential $V(x_3)$. The corresponding energy eigenvalue equation is 
\begin{align}
\left( \frac{-\hbar^2}{2 m}\partial_3^2 + V(x_3) \right)\chi(x_3)=E_0\chi(x_3) ,
\end{align}
with the ground state wavefunction $\chi(x_3)$  and the energy eigenvalue $E_0$. Then the field operator $\Psi_\alpha({\bm x},t)$ may be written as a product 
\begin{align}
\Psi_\alpha({\bm x},t) = \chi(x_3) \Phi_\alpha({\bm r},t), 
\end{align}
where the second quantized field operator $\Phi_\alpha({\bm r},t)$ describes two-dimensional electrons. We use the notation ${\bm r}=(x_1, x_2)$ throughout this work. The two-dimensional electron field operators $\Phi_\alpha({\bm r},t)$ and $\Phi_\beta^\dagger({\bm r},t)$ also satisfy the equal-time canonical anticommutation relation, 
\begin{align}
\left\{ \Phi_\alpha({\bm r},t), \Phi_\beta^\dagger({\bm r}',t)  \right\} = 
\delta_{\alpha\beta}\delta({\bm r}-{\bm r}') .
\label{ccr}
\end{align}
Using these field operators for the two-dimensional electrons and the wave function $\chi(x_3)$, we define the two-dimensional electron density operator $\rho_\alpha$ such that 
\begin{align}
\Psi_\alpha^\dagger({\bm x},t)  \Psi_\alpha({\bm x},t)
= 
\vert \chi(x_3) \vert^2  
\Phi_\alpha^\dagger({\bm r},t)  \Phi_\alpha({\bm r},t)
\equiv \vert \chi(x_3) \vert^2 \rho_\alpha({\bm r},t).
\end{align}
The two-dimensional retarded density-density response function is also defined as 
\begin{align}
{\mathcal D}_{\alpha\beta}({\bm r},t;{\bm r}',t')  = -i\theta(t-t') 
\left< \left[ 
\rho_\alpha({\bm r},t),  
\rho_\alpha({\bm r}',t')
\right] \right> .
\label{2DD}
\end{align}
Then the three-dimensional electron retarded density-density response function defined by Eq.  (\ref{3dimD}) may be expressed in terms of the function $\chi(x_3)$ and the two-dimensional electron retarded density-density response function: 
\begin{align}
{\mathcal D}^{(3D)}({\bm x},t;{\bm x}',t') = 
\vert \chi(x_3) \chi(x_3') \vert^2 
\sum_{\alpha\beta}
{\mathcal D}_{\alpha\beta}({\bm r},t;{\bm r}',t') .
\label{3D-2D}
\end{align}
Now the Poisson equation and the linear response equation for the 3-dimensional electron gas can be written as
\begin{align}
\nabla^2 A_0^{(1)} ({\bm x},t)= 4\pi e \epsilon^{-1}  
\vert \chi(x_3) \vert^2 \delta\left<\rho_\alpha({\bm r},t)\right>
\label{Poisson2D}
\end{align}
and 
\begin{align}
\vert \chi(x_3) \vert^2 \delta\left<\rho_\alpha({\bm r},t)\right> = 
\frac{- e}{\hbar}\int dt' \int d^3{\bm x}'  
\vert \chi(x_3) \chi(x_3') \vert^2 
\sum_{\alpha\beta}
{\mathcal D}_{\alpha\beta}({\bm r},t;{\bm r}',t') 
A_0^{(1)}({\bm x}',t') .
\label{SCLRA2D}
\end{align}
Equations (\ref{Poisson2D}) and (\ref{SCLRA2D}) constitute the basic equations of the SCLRA to determine the dispersion of longitudinal magnetoplasmons in the 2DES. By eliminating $\rho^{(1)}$ from them, we obtain a wave equation for the scalar potential with the retardation due to the electron dynamics. By eliminating $A_0^{(1)}$ from them, we obtain the plasmon wave equation. To obtain their dispersions, the retarded density-density response function of the electron gas must be calculated. 
From these equations one may examine quantitatively how the finite thickness of the two-dimensional electron system would affect the dynamics of the propagating collective mode.

\section{Retarded density-density response function for 2DEG in magnetic field}
In this section we calculate the retarded density-density response function of the 2DEG by expanding the electron field operator $\Phi_\alpha$ in terms of the Landau orbitals, which are eigenfunctions of the single-electron Schr{\" o}dinger equation
\begin{align}
\left\{ h_A (\partial) + \frac{g}{2}\mu_B B\sigma(\alpha)  \right\} 
v_{kn\alpha}({\bm r}) = E_{kn\alpha} v_{kn\alpha}({\bm r}) 
\label{Landaueq}
\end{align}
with the single-particle differential operator 
\begin{align}
h_A (\partial) = \frac{-\hbar^2}{2m}\left\{
\left( \partial_1 -\frac{ieB}{\hbar c} x_2 \right)^2 + \partial_2^2 \right\}- \mu .
\end{align}
Here $\mu$ is renormalized chemical potential for the two-dimensional system to include the $V(x_3)$-confining energy $E_0$ such that
\begin{align}
\mu = \mu_0 - E_0 .
\end{align}
The eigenfunctions of  Eq. (\ref{Landaueq}) are well-known Landau orbitals given by 
\begin{align}
v_{nk\alpha}(x_1,x_2) = \frac{e^{ikx_1}}{\sqrt{2\pi}}u_n(x_2-l^2k) 
\label{Landauorbital}
\end{align}
with the simple harmonic oscillator wave function
\begin{align}
u_n(x_2) = \frac{1}{\sqrt{2^n n! l \sqrt{\pi}}}\exp\left\{\frac{-x_2^2}{2l^2}\right\}
H_n\left( \frac{x_2}{l} \right) , 
\end{align}
where $l=\sqrt{c\hbar/eB}$ is the magnetic length and $H_n$ is the $n$th order Hermite polynomial. The energy eigenvalues are 
\begin{align}
E_{n\alpha}=\hbar\omega_c\left( n + \frac{1}{2} \right) + \frac{g}{2}\mu_B B\sigma(\alpha) -\mu ,
\end{align}
where $\omega_c=eB/mc$ is the cyclotron frequency. Now the second quantized electron field operator may be expanded in terms of the Landau orbitals given by Eq.  (\ref{Landauorbital}),  
\begin{align}
\Phi_\alpha({\bm r},t)= \int dk \sum_n C_{nk\alpha}(t) v_{nk\alpha}({\bm r}) .
\label{Phiexpansion}
\end{align}
The operator $C_{nk\alpha}$ and its hermitian conjugate satisfy the equal-time canonical anticommutation relation 
\begin{align}
\{C_{nk\alpha}(t), C_{n'k'\beta}^\dagger(t) \}=\delta_{\alpha\beta}\delta_{nn'}\delta(k-k') .
\label{cccr}
\end{align}
Writing the Hamiltonian in terms of these $C_{nk\alpha}$ and $C_{nk\alpha}^\dagger$ , we can readily find that their time-dependence is given as
\begin{align}
C_{nk\alpha}(t)=\exp\left( \frac{-i}{\hbar}E_{n\alpha}t \right) C_{nk\alpha} .
\label{ctime}
\end{align}
The density operator can be expanded as
\begin{align}
\rho_\alpha({\bm r},t) 
= \int dk \sum_{n=0}^\infty \int dk' \sum_{n'=0}^\infty 
\exp\left[ i\omega_c (n-n')t \right]
C_{nk\alpha}^\dag C_{n'k'\alpha}
v_{nk}^\ast({\bm r})  v_{n'k'}({\bm r})
\end{align}
The calculation of the expectation value for the commutator is given in Appendix A. The result is 
\begin{align}
\left<\left[ \rho_\alpha({\bm r},t), \rho_\beta({\bm r}',t') \right]\right> 
= \delta_{\alpha\beta} \sum_{n=0}^\infty \sum_{n'=0}^\infty e^{i\omega_c (n-n')(t-t')} 
\left\{
f\left( E_{n\alpha} \right) - f\left( E_{n'\alpha} \right)
\right\} {\mathcal M}_{n n'}({\bm r}, {\bm r}') 
\end{align}
where we have defined the Fermi distribution 
$
f\left( E_{n \alpha} \right) = \{1+\exp\beta( E_{n \alpha}-\mu )\}^{-1} 
$
and  
\begin{align}
{\mathcal M}_{n n'} ({\bm r},{\bm r}') 
= \int_{-\infty}^\infty dk \int_{-\infty}^\infty dk'
v_{nk}^\ast({\bm r})  v_{n'k'}({\bm r})
v_{n'k'}^\ast({\bm r}')  v_{n k}({\bm r}') 
\end{align}
with 
\begin{align}
{\mathcal M}_{n n'} ({\bm r},{\bm r}') 
= \frac{1}{(2\pi)^2} \int d^2{\bm k}  
\exp\left[ i{\bm k}({\bm r} - {\bm r}') \right] 
{\it\Lambda}_{n n'}({\bm k})
\end{align}
The function ${\it\Lambda}_{n n'}({\bm k})$ is calculated in Appendix A. We found 
\begin{align}
{\it\Lambda}_{n n'}({\bm k}) 
&= \frac{1}{2\pi l^2} \frac{n'!}{n!} 
X^{ n - n' } 
\exp\left(- X \right) 
\left\{ 
L_{n'}^{n - n'}
\left( X \right) 
\right\}^2 
\qquad ( n' \le n ) 
\end{align}
and
\begin{align}
{\it\Lambda}_{n n'}({\bm k}) 
&= \frac{1}{2\pi l^2} \frac{n!}{n'!} 
X^{ n' - n } 
\exp\left(- X \right) 
\left\{ 
L_n^{n' - n}
\left( X \right) 
\right\}^2 
\qquad ( n \le n' )
\end{align}
where $L_n^m$ is the associated Laguerre polynomial, and  
\begin{align}
X = \frac{l^2 k_2^2 + l^2 k_1^2}{2}= \frac{l^2 \vert {\bm k} \vert^2}{2} 
\end{align}
Using these ${\mathcal M}_{n n'} $ and ${\it\Lambda}_{n n'}$, we can calculate the Fourier transform of the retarded density-density response function defined by 
\begin{align}
{\mathcal D}_{\alpha\beta}({\bm r},t; {\bm r}',t') 
= \int \frac{d^2{\bm q}}{(2\pi)^2}\int \frac{d\omega}{2\pi} e^{i{\bm q}({\bm r}-{\bm r}')-i\omega (t-t')}
D_{\alpha\beta}({\bm q},\omega) .
\end{align}
Taking into account the time-dependence of the step function $\theta(t-t')$ in the frequency Fourier transform, we find  
\begin{align}
D_{\alpha\beta}({\bm q},\omega) = \frac{\delta_{\alpha\beta}}{2\pi l^2} 
\sum_{n=0}^\infty \sum_{n'=0}^\infty f(E_{n\alpha}) \Lambda_{nn'}(X) 
\qquad
\nonumber \\
\times \left\{
\frac{1}{\omega + \omega_c (n-n') +i\eta} - 
\frac{1}{\omega + \omega_c (n'-n) +i\eta}
\right\}.
\end{align}
This yields the real part of the density-density response function 
\begin{align}
{\rm Re} D_{\alpha\beta}({\bm q},\omega) = \frac{\delta_{\alpha\beta}}{2\pi l^2} 
\sum_{n=0}^\infty f(E_{n\alpha})  
\sum_{n'=0}^\infty \Lambda_{nn'}(X) 
\left\{
\frac{-2\omega_c (n-n')}{\omega^2 + \omega_c^2 (n-n')^2 } 
\right\} 
\qquad 
\end{align}
and the imaginary part 
\begin{align}
{\rm Im} D_{\alpha\beta}({\bm q},\omega) = 
\frac{\delta_{\alpha \beta}}{2 l^2}
\sum_{n=0}^\infty f(E_{n\alpha}) \sum_{n'=0}^\infty 
\qquad \qquad \qquad \qquad
\nonumber \\
\times 
\Lambda_{nn'}(X) 
\left\{
\delta\left( \omega - \omega_c (n-n') \right) - 
\delta\left( \omega - \omega_c (n'-n) \right)
\right\} .
\end{align}
In the next section we calculate the dispersion relations of the longitudinal plasmons in the long-wavelength limit. For that purpose, we need the expansion of the real part of $ D_{\alpha\beta}({\bm q},\omega)$ in powers of $X$. The expansion yields
\begin{align}
{\rm Re} D_{\alpha\beta}({\bm q},\omega) = 
\frac{\delta_{\alpha\beta}}{\pi l^2 \omega_c}
\left[
\left( \frac{\omega_c^2}{\omega^2-\omega_c^2} \right)
\left\{ \sum_{n=0}^\infty f(E_{n\alpha}) \right\} X\right]  + 
\nonumber \\
\frac{\delta_{\alpha\beta}}{\pi l^2 \omega_c}
\left[
\left( \frac{\omega_c^2}{\omega^2-4\omega_c^2} 
- \frac{\omega_c^2}{\omega^2-\omega_c^2} \right)
\left\{ \sum_{n=0}^\infty f(E_{n\alpha})(2n+1) \right\} X ^2 
\right] 
\label{Dexpansion}
\end{align}
which is exact up to the 2nd power of the variable $X$. We have not made any approximation except the expansion in power of $X$.

\section{SCLRA Equation for longitudinal magnetoplasmons}
In this section we derive the SCLRA equation to determine the dispersion relations of the longitudinal magnetoplasmons in 2DEG. If we define 
\begin{align}
D({\bm q}, \omega) = \sum_\alpha \sum_\beta {\rm Re}D_{\alpha\beta}({\bm q}, \omega) ,
\end{align}
then the Fourier transform of the linear response approximation gives 
\begin{align}
\delta \langle \rho^{(3D)}({\bm q},q_3 ,\omega) \rangle
=
\frac{-e}{2\pi \hbar} \sigma (q_3) D({\bm q},\omega) 
\nonumber \\
\times 
\int^{\infty}_{-\infty} dp_3 \sigma(-p_3) \Phi({\bm q},p_3 ,\omega) \;,
\label{response-formula-Fourier}
\end{align}
where $\sigma(q_3)$, $\delta \langle \rho^{(3D)}({\bm q},q_3 ,\omega) \rangle$, and $\Phi({\bm q}, q_3, \omega) $ are the Fourier transforms of ${|\chi(x_3)|^{2}}$, ${\delta \langle \rho^{(3D)}({\bm x},x_3 ,t) \rangle }$, and the scalar potential, respectively. 
Note that the $q_3$-dependence on the right-hand side of  Eq. (\ref{response-formula-Fourier}) appears only through $\sigma (q_3)$. 
Therefore, the function  
\begin{align}
\tilde\rho({\bm q},\omega) \equiv 
\frac{ \delta \langle \rho^{(3D)}({\bm q},q_3 ,\omega) \rangle}
{\sigma (q_3)} 
\label{rhotilde}
\end{align}
is independent of $q_3$.
On the other hand the Fourier transform of Poisson's equation (\ref{Poisson}) gives
\begin{align}
\Phi({\bm q},q_3 ,\omega) = 
\frac{ -4 \pi e \delta \langle \rho^{(3D)}({\bm q},q_3 ,\omega) \rangle}
{\epsilon \big[ q^2 + q^{2}_{3} \big]}\;,
\label{Poisson-Fourier}
\end{align}
which can be written as
\begin{align}
\Phi({\bm q},q_3 ,\omega) = 
\frac{ -4 \pi e \sigma (q_3)}
{\epsilon \big[ q^2 + q^{2}_{3} \big]}
\tilde\rho({\bm q},\omega) \;.
\label{Poisson-Fourier2}
\end{align}
Substituting (\ref{Poisson-Fourier2}) into (\ref{response-formula-Fourier}), we obtain
the SCLRA equation
\begin{align}
\left\{
1-\frac{2e^2}{\hbar \epsilon}\Gamma(q)D({\bm q},\omega)
\right\} 
\tilde{\rho}({\bm q},\omega)=0 \;,
\label{SCLRA2}
\end{align}
where we have defined 
\begin{align}
\Gamma(q)
=
\int^{\infty}_{-\infty} dp_3 
\frac{\sigma(-p_3) \sigma(p_3)}{|{\bm q}|^2+p_3^2} \;.
\label{Gamma}
\end{align}
If there exists collective density fluctuation of the electrons such as longitudinal plasmon mode, then we have a non-vanishing $\tilde{\rho}$ in Eq. (\ref{SCLRA2}). This leads to the following equation to determine the dispersion of the longitudinal plasmon collective mode: 
\begin{align}
1 - \frac{2 e^2}{\hbar\epsilon} \Gamma(q) D({\bm q},\omega)=0 \;.
\label{dis-1}
\end{align}
For the wavefunction $\chi(x_3)$ We assume a Gaussian model 
\cite{FHMT2007}
\begin{align}
|\chi(x_3)|^2= \frac{a}{\sqrt{2\pi}}\exp\left( \frac{-a^2 x_3^2}{2}  \right) ,
\end{align}
then 
\begin{align}
\Gamma(q)
=2\sqrt{\pi}{q^{-1}} \exp(q^2 / a^2){\rm Erfc}(q/a) 
\end{align}
Here ${\rm Erfc}(x)$ is the complementary error function 
${
{\rm Erfc}(x) \equiv \int^{\infty}_{x} ds e^{- s^2}
}$.
For small $q/a$ this $\Gamma$ can be expanded as
\begin{align}
\Gamma(q)
=
\frac{\pi}{q} 
\bigg[ 
1- \frac{2}{\sqrt{\pi}} \left(\frac{q}{a}\right) 
+ \left(\frac{q}{a}\right)^2 
- \frac{4}{3\sqrt{\pi}} \left(\frac{q}{a}\right)^3 
\nonumber \\ 
+ \frac{1}{2} \left(\frac{q}{a}\right)^4 
- \frac{8}{15\sqrt{\pi}}\left(\frac{q}{a}\right)^5 
+ \cdot \cdot \cdot \,\,
\bigg] .
\label{gamma-large-a}
\end{align}
The value of the parameter $a$ is given by the thickness of the 2DEG layer. Since we have adopted the Gaussian model for $|\chi(x_3)|^2$, in order to define the thickness $d$ it would be reasonable to assume $\int_{-d/2}^{d/2} |\chi(x_3)|^2 dx_3 = 0.99$, which yields $a= 5/d$ \cite{FHMT2007}.

\section{Dispersion relations of longitudinal magnetoplasmons}
Substituting the expansions given by  Eqs. (\ref{Dexpansion}) and (\ref{gamma-large-a}) into  (\ref{dis-1}), we can straightforwardly derive the dispersion relations of the magnetoplasmons. The only approximation used in the calculation is the expansion in power of $lq$. We consider up to the fourth power of $q$. Then we obtain the following two magnetoplasmon modes: 
\begin{align}
\frac{\omega^2}{\omega_c^2} &= 1+A_1\kappa lq 
- \frac{2A_1 \kappa}{\sqrt{\pi} \lambda}(lq)^2 
+\left( \frac{A_1 \kappa}{\lambda^2}-\frac{A_2 \kappa}{2} \right) (lq)^3 
\nonumber \\ 
&\qquad \qquad +\left( 
\frac{-4A_1 \kappa}{3\sqrt{\pi} \lambda^3} 
+ \frac{A_2 \kappa}{\sqrt{\pi} \lambda} 
- \frac{A_1 A_2 \kappa^2}{6} 
\right)(lq)^4 
\nonumber \\
&\equiv \Omega_1^2(lq) 
\label{dispersion1}
\end{align}
and
\begin{align}
\frac{\omega^2}{\omega_c^2}= 4 + 
\frac{A_2 \kappa}{2}(lq)^3
+\left(\frac{A_1 A_2 \kappa^2}{6}
-\frac{A_2 \kappa}{\sqrt{\pi} \lambda}
\right)(lq)^4
\equiv \Omega_2^2(lq) .
\label{dispersion2}
\end{align}
Here we have defined 
\begin{align}
A_1 = \sum_\alpha \sum_{n=0}^\infty f(E_{n\alpha}) 
\label{A1}
\end{align}
and
\begin{align}
A_2 = \sum_\alpha \sum_{n=0}^\infty f(E_{n\alpha})(2n+1) .
\label{A2}
\end{align} 
These quantities $A_1$ and $A_2$ appear frequently in the transport theory of a many-electron system in a quantizing magnetic field \cite{Greene69}. 
Physically this $A_2$ is proportional to the internal energy of the 2DES due to the cyclotron motion of the electrons. 
We have also introduced the dimensionless parameters $\kappa$ and $\lambda$ defined as 
\begin{align}
\kappa \equiv \frac{e^2}{\epsilon l}\frac{1}{\hbar \omega_c}
\qquad {\rm and} \qquad 
\frac{1}{\lambda} \equiv \frac{d}{5l}.
\label{kappalambda}
\end{align}
Roughly speaking, the parameter $\kappa$ is the ratio of the average Coulomb interaction energy to the cyclotron energy, and $\lambda$ is the ratio of the magnetic length (times 5) to the thickness of the 2DEG.\\
\indent
If the terms with $\lambda$ in the dispersion relations (\ref{dispersion1}) and (\ref{dispersion2}) are neglected, i.e., if the zero layer thickness limit is taken, then the dispersion relations agree with those obtained by Horing and Yildiz \cite{Horing76} except the fact that the coupling between spin and magnetic field, and the zero-point energy contribution to the cyclotron energy spectrum in the coefficient $A_2$ defined by  Eq. (\ref{A2}) are neglected in their calculation.

The dispersion (\ref{dispersion1}) agrees with the well-known result. Taking up to the first order terms in the $lq$-expansion of Eq. (49) and substituting (\ref{A1}) and (\ref{kappalambda}) into (\ref{dispersion1}), we find 
\begin{align}
{\omega^2} = {\omega_c^2} + \frac{2\pi e^2 n_{\rm exp}}{m \epsilon} q \;.
\label{dispersion3}
\end{align}
In $B\rightarrow 0$ limit this gives the well known two-dimensional plasmon dispersion \cite{Stern1967}.
Eq. (\ref{dispersion3}) corresponds to one of the dispersions obtained by Bernstein \cite{Bernstein}, who solved the coupled Maxwell-Boltzmann equations in the linear approximation. The difference is that Bernstein's calculation is fully classical, while our calculation is based on the rigorous quantum many-body theory. The quantum effects can be most clearly seen in the expansion (\ref{dispersion1}) and (\ref{dispersion2}), where the Fermi distribution function appears explicitly in the coefficients $A_1$ and $A_2$. 

Recently Gudmundsson {\it et al.} \cite{Gudmundsson1995} thoroughly investigated both theoretically and experimentally the magnetoplasmon dispersion in quantum dots and showed the Bernstein modes are also found in such low dimensional systems. This seems to be due to the fact that the Bernstein modes are essentially classical effect. They pointed out that theoretically it cannot be cleanly observed in exact numerical diagonalization for few particles. Since the quantum field theoretical calculation is also applicable to a system of few electrons, it would be an interesting problem to apply the present theoretical method to quantum dots. 

To end this section we would like to make a brief remark on the Coulomb interaction between electrons in the SCLRA. The effect of the Coulomb interaction on the dynamics of the electrons is taken into account by solving simultaneously the Poisson equation and the linear response equation. This yields the collective modes of the electron density caused by the Coulomb interaction. On the other hand, the effect of the Coulomb interaction on the many-electron ground state is not explicitly calculated, as it has been well-established that such effect may change the effective mass but not the Fermi liquid nature.

\section{Comparison with experiment}
In this section we examine the validity of the  dispersion of the magnetoplasmon (\ref{dispersion1}) and (\ref{dispersion2}) using the experimental data given by Batke et al. \cite{Batke86}. 
Let us first look into $A_1$ and $A_2$, which contains the Fermi distribution function that manifests the explicit quantum statistical nature of the 2DES, i.e., Pauli's exclusion principle expressed by the equal-time anticommutation relation. It should be noted that the temperature dependence of the retarded density-density response function comes into only through the Fermi distribution.  The sum of the Fermi distribution in $A_1$ simply yields the electron number density, 
\begin{align}
\frac{eB}{hc}\sum_\alpha\sum_{n=0}^\infty f(E_{n\alpha}) = n_{\rm exp}. 
\end{align}
Then, we find 
\begin{align}
A_1 = \frac{hc}{eB} n_{\rm exp}.
\label{A1-2}
\end{align} 
Similarly $A_2$ may be expressed in terms of the internal energy of the 2DES. However, in the experiment, the sample is immersed into a heat reservoir and the temperature is fixed. 
On the other hand, the chemical potential in the Fermi distribution in $A_2$ must be expressed as a function of the electron number density $n_{\rm exp}$, the temperature, and other parameters relevant to a given experimental situation. However, if we consider the zero-temperature case, we may straightforwardly calculate this $A_2$ for a given electron number density. 
At very low temperatures, the summation in the definition of $A_2$ given by (\ref{A2}) can be approximated as 
$
\sum_\alpha \sum_{n=0}^\infty f(E_{n\alpha})n 
\simeq M(M+1)
$, 
where $M$ is the highest Landau level occupied by electrons at zero-temperature. 
The same approximation applied to the coefficient $A_1$ yields 
$2M+2=A_1$. Therefore, we obtain 
\begin{align}
A_2 = 2 \sum_\alpha \sum_{n=0}^\infty f(E_{n\alpha})n + A_1 
\simeq {A_1^2}/{2} \; .
\label{A2-2}
\end{align}
Using eqs. (\ref{A1-2}) and (\ref{A2-2}), the dispersion given by (\ref{dispersion1}) and (\ref{dispersion1}) can be expressed as a function of $B$ and $q$. As the plasmon frequency  $\nu=\omega/2\pi c$ (cm$^{-1}$) is plotted as a function of $B$ for the given values of the wave vector $q$ in \cite{Batke86}, let us define $\nu_i$ such that  
\begin{align}
\nu_i (B,q)\equiv \frac{\omega_c}{2\pi c}\Omega_i(lq) \qquad (i=1,2) 
\label{nui}
\end{align}
We use the parameters given in \cite{Batke86}. 
The electron number density $n_s$ and thickness $d$ of the Al$_x$Ga$_{1-x}$As-GaAs heterostructure 2DES sample used in the measurement are $n_s = 6.7\times 10^{11}$ (cm$^{-2}$) and $d = 8 \times10^{-6}$ (cm). The effective mass $m$ and the dielectric constant $\epsilon$ given in \cite{Batke86} are $m = 0.071m_0$, where $m_0$ is the electron rest mass, and $\epsilon = 12 $, respectively. 
With these parameters we plot $\nu_1$ and $\nu_2$ as functions of $B$ for the wave vector $q_1=0.72\times 10^5$ (cm$^{-1}$) in Fig. 1 and for $q_2=1.44\times 10^5$ (cm$^{-1}$) in Fig. 2. The measured plasmon frequencies given in Fig. 7 of \cite{Batke86} are also shown in the graphs. The theoretical dispersion for the two different wave vectors are plotted together in Fig. 3. 

In Fig. 1 the theoretical curves agree with the measured frequency points fairly well for $B > B_c(q_1)$ with $B_c (q_1) \simeq 1$ (T), but considerable deviation is observed for $B<B_c(q_1)$. Similarly in Fig. 2 the theoretical curves agree with the measured frequency points fairly well for $B > B_c(q_2)$ with $B_c (q_2) \simeq 2$ (T), but considerable deviation is observed for $B<B_c(q_2)$ (T). This deviation seems to be caused by the expansion in power of $lq$, which is the only approximation used in the calculation of the dispersion given by (\ref{dispersion1}) and (\ref{dispersion2}). Although these two data may not be sufficient to infer the explicit $q$-dependence of $B_c$, the values for $B_c (q_1)$ and $B_c (q_2)$ found here may give certain criteria to apply the theoretical results (\ref{dispersion1}) and (\ref{dispersion2}) to investigate experiments.

\section{Concluding remarks}
We have obtained explicit finite layer thickness dependence of the dispersion relations of the longitudinal plasmons in a 2DEG in the presence of a quantizing magnetic field. 
The analytical calculation presented in this article is a rigorous extension of the previous works by Chiu and Quinn \cite{Chiu74} and by Horing and Yildiz \cite{{Horing73},{Horing76}} to finite layer thickness. The quantum field theoretical method of calculating the retarded density-density response function and dispersion relations presented in this work is much simpler than the previous work. The same theoretical method can be easily applied to the transverse plasmon, which is directly related to optical properties. This simplicity seems to be of great advantage in further applications to other many-electron systems such as the zero-mass Dirac field model for graphene 2DES.

We have also examined the validity of the $lq$ expansion used in the derivation of the dispersion relations (\ref{dispersion1}) and (\ref{dispersion2}) by comparing theoretical $B$ dependence with the experimental data given in \cite{Batke86} and found good agreement for $B>B_c(q)$. This result is of practical significance because the $lq$ expansion in the analytical calculation of the plasmon dispersion seems to be unavoidable and because most experimental measurements of magnetoplasmon dispersion are carried out by varying $B$ for fixed values of the wave vector.

Lastly we would like to comment on the electron reservoir model (ERM) of a 2DES under quantizing magnetic field. In 2004 Holland et al. \cite{Holland04} found plateaus in the coefficient of $q$ in the magnetoplasmon dispersion as a function of the filling factor in GaAs quantum well 2DES. They remarked that the phenomenon bears a striking resemblance to the quantum Hall effect (QHE) \cite{Klitzing}. Their remark was theoretically confirmed by Toyoda et al. \cite{Toyoda2008}, who showed the plateaus are perfectly explained by the ERM 
\cite{{Baraff},{TGT-PL},{TGT-Physica},{Toyoda-MPLB},{Toyoda-Zhang}}. In the ERM, the Fermi distribution function in the definitions of $A_1$ and $A_2$ should be regarded as a function of $T$, $B$, and the chemical potential $\mu$. Then $A_1$ shows plateaus as a function of $B$ similarly to the QHE. At the same time, the temperature dependence can be explicitly calculated. This may explain the slight deviation of the theoretical dispersion curves from the experiments in Fig. 1 and Fig. 2. The examination of the dispersion given by (\ref{dispersion1}) and  (\ref{dispersion2}) on the basis of the ERM is left for the future study.

\newpage
\appendix
\section{Calculation of $\Lambda_{nn'}$}
In order to simplify calculations let's define
\begin{align}
1 \equiv \left( n,k,\alpha \right), \quad 
2 \equiv \left( n',k',\alpha \right), \quad 
3 \equiv \left( n'',k'',\beta \right), \quad 
4 \equiv \left( n''',k''',\beta \right)
\end{align}
and $\delta(i,j)$ such that
\begin{align}
\delta(2,3) = \delta(k'-k'')\delta_{n',n''}\delta_{\alpha \beta}
\end{align}
Then the commutator appears in the response function can be computed as
\begin{align}
\left[ C_1^\dag C_2, C_3^\dag C_4 \right]
= \delta(2,3) C_1^\dag C_4 
- \delta(4,1) C_3^\dag C_2
\end{align}
By introducing
\begin{align}
\int_1 \equiv \int dk \sum_{n=0}^\infty \;, \quad
\int_2 \equiv \int dk' \sum_{n'=0}^\infty \;, \quad ...
\end{align}
the expectation value for the commutator can be written as
\begin{align}
\left< 
\left[ \rho_\alpha({\bm r},t), \rho_\beta({\bm r}'',t'') \right] \right>  
&= \int_1 \int_2 \int_3 \int_4
\exp i\omega_c \left[ (n-n')t + (n''-n''')t'' \right] 
\nonumber \\ 
&\times
v_{nk}^\ast({\bm r})  v_{n'k'}({\bm r})
v_{n''k''}^\ast({\bm r}'')  v_{n'''k'''}({\bm r}'') 
\nonumber \\ 
&\times 
\left\{
\delta(2,3)\delta(4,1) \left< C_1^\dag C_1 \right> 
- \delta(4,1)\delta(2,3) \left< C_2^\dag C_2 \right>
\right\}
\end{align}
The expectation value $\langle C_j^\dag C_j \rangle$ is given in terms of the Fermi distribution as
\begin{align}
\langle C_1^\dag C_1 \rangle = 
\langle C_{n k \alpha}^\dag C_{n k \alpha} \rangle 
= \frac{1}{1+\exp\beta\left[ E_{n \alpha}-\mu \right]} 
= f\left( E_{n \alpha} \right)
\end{align}
and
\begin{align}
\langle C_2^\dag C_2 \rangle = 
\langle C_{n' k' \alpha}^\dag C_{n' k' \alpha} \rangle 
= \frac{1}{1+\exp\beta\left[ E_{n' \alpha}-\mu \right]} 
= f\left( E_{n' \alpha} \right)
\end{align}
which are independent of $k$ and $k'$.
Thus the expectation value for the commutator can be written as
\begin{align}
\left<\left[ \rho_\alpha({\bm r},t), \rho_\beta({\bm r}',t') \right]\right> 
= \delta_{\alpha\beta} \sum_{n=0}^\infty \sum_{n'=0}^\infty e^{i\omega_c (n-n')(t-t')} 
\left\{
f\left( E_{n\alpha} \right) - f\left( E_{n'\alpha} \right)
\right\} {\mathcal M}_{n n'}({\bm r}, {\bm r}') 
\end{align}
where we have defined 
\begin{align}
{\mathcal M}_{n n'} ({\bm r},{\bm r}') 
= \int_{-\infty}^\infty dk \int_{-\infty}^\infty dk'
v_{nk}^\ast({\bm r})  v_{n'k'}({\bm r})
v_{n'k'}^\ast({\bm r}')  v_{n k}({\bm r}') 
\end{align}
Let us introduce new dimensionless integral variables, 
\begin{align}
\zeta \equiv \frac{x_2}{l}, \quad \zeta' \equiv \frac{x'_2}{l}, \quad 
\eta \equiv lk_1, \quad \xi \equiv lk_2 .
\end{align}
and define  
\begin{align}
W_n(s) = \frac{i^n }{\sqrt{2\pi}\sqrt{2^n n! l \sqrt{\pi}}}
\exp\left( \frac{-s^2}{2} \right) H_n(s) 
\label{FT-Wn}
\end{align}
Then  ${\mathcal M}_{n n'}$ can be written as 
\begin{align}
{\mathcal M}_{n n'} ({\bm r},{\bm r}') 
&= \int_{-\infty}^\infty dk_1 \frac{1}{(2\pi)^2}\exp\left[ ik_1(x_1 - x'_1) \right] 
M_{n n'}( \zeta, \zeta')
\label{M}
\end{align}
with $M_{n n'}$ defined as
\begin{align}
M_{n n'}( \zeta, \zeta') 
&= \frac{2\pi}{l} 
\int_{-\infty}^\infty d\xi \exp\left[i\xi(\zeta - \zeta') \right] 
\int_{-\infty}^\infty ds_1 
\exp(i\eta s_1) W_n(s_1) W_{n'}(\xi - s_1) 
\nonumber \\
&\times 
\int_{-\infty}^\infty ds_2   
\exp(i\eta s_2) W_n(s_2)  W_{n'}(-\xi - s_2) 
\end{align}
Hence the computation of ${\mathcal M}_{n n'}$ reduces to the integral
\begin{align}
I_{n n'}(\xi,\eta) \equiv  
\int_{-\infty}^\infty ds \exp(i\eta s) W_n(s) W_{n'}(\xi - s) 
\label{Inn1def}
\end{align}
which gives
\begin{align}
M_{n n'}( \zeta, \zeta') 
= \frac{2\pi}{l} 
\int_{-\infty}^\infty d\xi \exp\left[i\xi(\zeta - \zeta') \right] 
I_{n n'}(\xi,\eta) I_{n n'}( - \xi,\eta)
\label{Mnn}
\end{align}
The integral $I_{n n'}(\xi,\eta)$ can be written as 
\begin{align}
I_{n n'}(\xi,\eta) = C_{n n'} 
\exp\left\{ \left(\frac{\xi+i\eta}{2} \right)^2 - \frac{\xi^2}{2} \right\} 
J_{n n'}(\xi,\eta) 
\label{InnJnn}
\end{align}
with
\begin{align}
C_{n n'} = \frac{1}{2\pi}
\frac{i^n }{\sqrt{2^n n! l \sqrt{\pi}}} 
\frac{i^{n'}}{\sqrt{2^{n'} n'! l \sqrt{\pi}}} 
\label{Cnn}
\end{align}
and  
\begin{align}
J_{n n'}(\xi,\eta) 
= 
\int_{-\infty}^\infty ds \exp\left[ 
-\left( s- \frac{\xi + i\eta}{2} \right)^2\right]
H_n(s) H_{n'}(\xi-s) 
\end{align}
This integral can be readily evaluated \cite{GR}. 
For $n' \le n$ we find 
\begin{align}
J_{n n'}(\xi,\eta) 
= (-)^{n'} 2^n \sqrt{\pi} \{ n'! \}
\left( \frac{\xi + i\eta}{2} \right)^{n -n'} 
L_{n'}^{n-n'} \left( \frac{\xi^2 + \eta^2}{2} \right) 
\label{J1n'<n}
\end{align}
For $n \le n'$ we find 
\begin{align}
J_{n n'}(\xi,\eta) = (-)^{n'} 2^{n'} \sqrt{\pi} \{ n! \} 
(-)^{n'-n} 
\left( \frac{\xi - i\eta}{2} \right)^{n'-n} 
L_{n}^{n'-n} \left( \frac{\xi^2 + \eta^2}{2} \right) 
\label{J1n<n'}
\end{align}
These two results yield  
\begin{align}
&I_{n n'}(\xi,\eta) I_{n n'}(-\xi,\eta) 
\nonumber \\
&= \frac{1}{(2\pi)^2 l^2} \frac{N_<!}{N_>!} 
\left( \frac{\xi^2 + \eta^2}{2} \right)^{ N_> \; - \; N_< }
\exp\left\{- \left( \frac{\xi^2+\eta^2}{2} \right) \right\} 
\left\{ 
L_{N_<}^{N_> \; - \; N_<}
\left( \frac{\xi^2 + \eta^2}{2} \right) 
\right\}^2 
\end{align}
where $N_>$ is the larger of $n$ and $n'$, and $N_<$ is the smaller. Now we define 
\begin{align}
X = \frac{\xi^2+\eta^2}{2}
= \frac{l^2 k_2^2 + l^2 k_1^2}{2}= \frac{l^2 \vert {\bm k} \vert^2}{2} 
\end{align}
and use (\ref{M}) and (\ref{Mnn}) to obtain  
\begin{align}
{\mathcal M}_{n n'} ({\bm r},{\bm r}') 
= \frac{1}{(2\pi)^2} \int d^2{\bm k}  
\exp\left[ i{\bm k}({\bm r} - {\bm r}') \right] 
{\it\Lambda}_{n n'}({\bm k})
\end{align}
where the function ${\it\Lambda}_{n n'}({\bm k})$ is defined as
\begin{align}
{\it\Lambda}_{n n'}({\bm k}) 
&= \frac{1}{2\pi l^2} \frac{n'!}{n!} 
X^{ n - n' } 
\exp\left(- X \right) 
\left\{ 
L_{n'}^{n - n'}
\left( X \right) 
\right\}^2 
\qquad ( n' \le n ) 
\end{align}
and
\begin{align}
{\it\Lambda}_{n n'}({\bm k}) 
&= \frac{1}{2\pi l^2} \frac{n!}{n'!} 
X^{ n' - n } 
\exp\left(- X \right) 
\left\{ 
L_n^{n' - n}
\left( X \right) 
\right\}^2 
\qquad ( n \le n' )
\end{align}

\newpage
\noindent
{\bf Figure captions}\\
\\
Fig. 1\\
Dispersion $\nu_1(B,q) $ and $\nu_2(B,q) $ given by Eq. (\ref{nui}) are plotted as a function $B$ for the wave vector $q=q_1= 0.72\times 10^5$ (cm$^{-1}$). The measured magnetoplasmon frequencies given in Fig. 7 in \cite{Batke86} are shown by by small black circles. The two dotted lines show the cyclotron frequency $\omega_c/2\pi c$ and $2\omega_c/2\pi c$, respectively.
\\
\\
\\
\\
\\
Fig. 2\\
Dispersion $\nu_1(B,q) $ and $\nu_2(B,q) $ given by Eq. (\ref{nui}) are plotted as a function $B$ for the wave vector $q=q_2= 1.44\times 10^5$ (cm$^{-1}$). The measured magnetoplasmon frequencies given in Fig. 7 in \cite{Batke86} are shown by small black triangles. The two dotted lines show the cyclotron frequency $\omega_c/2\pi c$ and $2\omega_c/2\pi c$, respectively.
\\
\\
\\
\\
\\
Fig. 3\\
The solid curves are the dispersion $\nu_1(B,q_1) $ and $\nu_2(B,q_1)$ given by Eq. 
(\ref{nui}) for the wave vectors  $q_1= 0.72\times 10^5$ (cm$^{-1}$). The dashed curves are the dispersion 
$\nu_1(B,q_2) $, and $\nu_2(B,q_2) $, given by Eq. (\ref{nui}) for the wave vectors 
$q_2= 1.44\times 10^5$ (cm$^{-1}$).

\end{document}